\documentclass[11pt]{article}

\usepackage{graphicx, amssymb, latexsym, amsfonts, amsmath, lscape, amscd,
amsthm, color, epsfig, mathrsfs, tikz, enumerate, lipsum}
\usepackage{fullpage}
\usepackage{graphicx}
\usepackage{epstopdf}
\usepackage{float}
\usepackage{caption}
\usepackage{pstricks}
\usepackage{subfig}

\begin{document}

\title{{\bf Resource allocation determines alternate cell fate in Bistable Genetic Switch }}
\author{
{\sc Priya Chakraborty}$^{1}$, {\sc Sayantari Ghosh}$\,^2$,\\
\mbox{}\\
{ Department of Physics, National Institute of Technology,
Durgapur, India.}\\
{ Email:  pc.20ph1104@phd.nitdgp.ac.in \textsuperscript{1} ;   sayantari.ghosh@phy.nitdgp.ac.in \textsuperscript{2} }\\}
%


\maketitle
\section*{Abstract}\fontsize{12}{1}
\indent Living cells need a constant availability of certain resources to have a sustained gene expression process. Limited availability of cellular resources  for gene expression, like ribosomes, along with a variation of resource affinity, significantly modifies the system dynamics. Factors like the variation in rate of binding, or variation in efficiency of the recruited resource have the potential to affect crucial dynamical phenomena like cell fate determination. In this paper, we have taken a very important motif, a bistable genetic toggle switch, and explored the effect of resource imbalance in this circuit in terms of the bifurcations taking place. We show that initial asymmetric biasing to resource via resource affinity or gene copy number, significantly modifies the cell fate transition, both in pitchfork and saddle node type bifurcation. Our study establishes that in a limited resource environment, controlled resource allocation can be an important factor for robust functioning of the synthetic or cellular genetic switches.

\noindent{\it Keywords}: Gene regulation, Resource allocation, Cell fate decision, Pitchfork bifurcation, Saddle node bifurcation, Genetic toggle switch. 
\section{Introduction}
Proteins that govern all functionalities in a living cell are produced by two major steps: transcription and translation, which indeed are a combination of several intermediate process. In order to understand intriguing cellular  processes like, cellular decision making, or to operate a synthetic circuit inside cell, detailed mathematical study of the cellular or synthetic gene regulatory dynamics is of prime concern. Extreme non-linearity inside cellular systems and the nature of coupling of one ongoing process with other makes this understanding of the dynamics extremely difficult. Luckily some re-occurring genetic pattern called motif, can be found in living organisms which performs some typical tasks for the cell. Scientists are focusing largely to explore this repeatedly occurring motifs to understand the cell dynamics as well as cellular decision making from last few decades. 
Genetic toggle is one of the most detailed studied biological motifs \cite{gardner2000construction,strasser2012stability,tian2004bistability}, where two gene (lacI and tetR) mutually repress each other's expression. Dynamics of this genetic motif is usually studied from the perspective of saddle-node bifurcation. Here in case of saddle node bifurcation of toggle, the system undergo a transition to bistable state from a monostable one, and after a specific range of parameters again the system become monostable. Thus, a specific region of bistability separates two otherwise monostable region in the phase plane. The system can also undergo for a pitchfork bifurcation for some specific symmetry as studied in some recent works \cite{bose2019bifurcation}. In case of pitchfork type bifurcation, a monostable region converts to a bistable one at a transition point and remains bistable for rest. Bistabilty introduces some irreversibility in biological systems, that once the system attains its steady state, it retains its same steady state even on application of some output perturbations in small scale. Thus, in case of cell fate differentiation and cellular decision making,  genetic toggle motif is taken as a canonical circuit for understanding \cite{balazsi2011cellular}.
\\From its early invention as a synthetic circuit by Gardner et al.\cite{gardner2000construction} different approaches for a robust controlling of genetic toggle is proposed. Some of them involves real time feedback control\cite{lugagne2017balancing}, auto regulation, noise\cite{wang2007noise}, addition of some diffusible molecule like isopropyl-$\beta$-D-thiogalactopyranoside (IPTG) and anhydrotetracycline (aTc) to control the promoter activity as well. The search for novel control parameter is still going on, and in some recent study it is established that  the limited availability of cellular ingredients, serving in the process of protein production can act as robust parameter in cell dynamics. 
\\ In the intermediate steps of protein production, majorly transcription and translation, the genes collect resources for successful completion of its expression from the cell. RNAP, transcription factor (TF), ribosome, degradation machinery etc. are various resources the cell supply to the synthetic genetic circuit implemented in it, or uses for its endogenous gene functionality. It is experimentally verified that the cell does not contain these resources infinitely; even depending upon the mode of operation, availability of resource in different cells vary significantly. In translation process, ribosome is considered to be the most important resource the gene circuit collects from the cell. Though the presence of this essential cellular resource in all the prokaryotes and eukaryotes is a fact,
the amount of free ribosome in different living being is different, even for the different cell having different functionality differs in availability of free ribosomes. Like the pancreatic cells, in eukaryotes, dedicated for most of the protein production than other cells, contains unusually high number of ribosomes, but in contrary Smooth endoplasmic reticulum (SER), does not
have ribosomes on its surface, and thus does not participate in protein production. This limited availability of the essential translational resource inside cell significantly affects the ongoing dynamics. In a low protein activity, or for a demand of low resource for the implemented synthetic construct, this limited availability may not affect, but in higher protein activity, or for a larger resource demand for the synthetic construct, unprecedented resource competition comes into picture. Different experimental and theoretical studies establishes that ribosome limitation significantly modifies the circuit dynamics and in larger scale the system chooses for favourable state. In a recent study, it is established that protein production curve can be largely modified in terms of sensitivity and amplification by controlling the ribosome availability and its distribution in system \cite{chakraborty2021emergentb}. Due to nature of coupling of different ongoing process, the competition in outer motifs, which seems apparently not important in study of motif interest significantly modifies the dynamical behaviour, even can ruin the entire system, resulting in some emergent response in output \cite{chakraborty2021emergent}. Not only limited to ribosomes, this competition can arise for RNAP, gene copy numbers, degradation machinery's and many more. Yuriy  Mileiko and team in a recent study have shown that gene copy number variation brings a significant change in the dynamics of some well known motifs \cite{mileyko2008small}. The effect of decoy binding is also established by some recent papers as well\cite{lee2012regulatory, jayanthi2013retroactivity}. This trend of exploring the consequences of resource availability and distribution in cell dynamics is new, and scientists are getting exciting novel mechanisms of system modelling from it.
\\Cellular decision-making, environmental sensing and cell to cell communication are three key processes underlying pattern formation and development in microscopic to the complex organisms. 
From a theoretical point of view though the cellular decision making seems to be reversible, in practice most of the time this is irreversible due to some secondary effects arising from the process. The biologically programmed cell death (apoptosis) or in response to the injury, the cell death (lysis) are most promising reason here. Several approaches in determining cell fate like feedback controlled regulation, cell size, growth rate dependency \cite{tan2009emergent,klumpp2009growth} is established in recent past. The effect of noise, especially when a group of cell is participating, statistics predicts the most probable as shown in \cite{balazsi2011cellular}. Though the presence of these various efforts of determining cell fate transition, the effect of limited availability of resource or asymmetry in resource affinity inside cell, in the cell fate dynamics is not explored much. This asymmetry serving as initial biasness has the capacity to pre-pattern the cell fate.
\\In this paper, we have taken a simple genetic toggle motif and considered along with the mutual repression to each other, both the genes are collecting resource, (say ribosome here) from the same pool with different affinity. We also explore the condition for variability in gene copy numbers for the pitchfork type bifurcation here. Some of the major findings of this paper are
\begin{itemize}
    \item Variability in resource affinity as well as gene copy number introduces asymmetry in pitchfork bifurcation and pre-patterns the cell fate. Greater the asymmetry, more the system pre-patterns itself in determining alternate cell fate.
    \item Asymmetry in resource affinity regulates the range of bistability, thus robustness of the switch in saddle node bifurcation.
    \item Total availability of resource regulates the point of bifurcation and region of interest in the system.
\end{itemize}

\section{Model Formulation}\label{model}
\begin{figure}
    \centering
    \includegraphics[width=16cm]{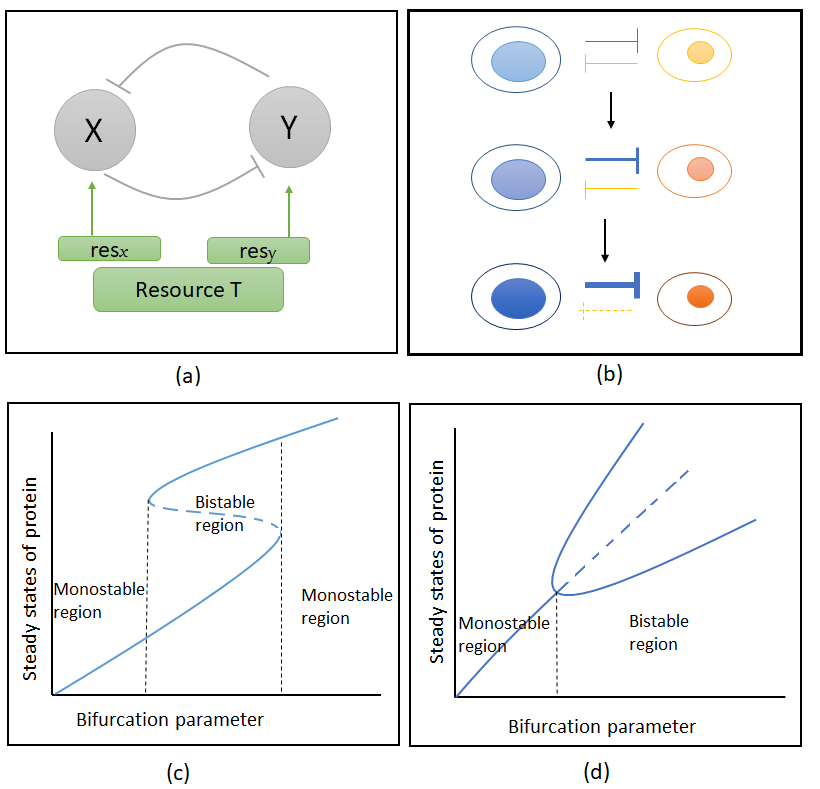}
    \caption{Model motif: (a). Genetic toggle, two protein $X$ and $Y$ mutually repress each others promoter activity. Hammer head symbol represents the repression here. Also both the participants of toggle collects resource from same pool $T$ with the affinities $res_x$ and $res_y$ respectively. (b). Schematic diagram of inter cellular competition determining cell fate. Strength of repression drives the cell to a particular fate. (c). Saddle node bifurcation in genetic toggle. A bistable region separates two monostable region in phase space. (d). Pitchfork bifurcation in genetic toggle. Monostable region switches to a bistable region at bifurcation point.}
    \label{motif}
\end{figure}
Let us consider, $X$ and $Y$ are the two proteins, repressing each others promoter activity, forming a toggle switch as shown in Fig. \ref{motif}.  The hammerhead symbols represent the repression here. We also consider both this proteins are collecting resource from the same pool for their expression. We particularly focus on the pool of ribosome here, the essential resource for translation process, for the transcriptional complex to be translated as protein product. 
\\Moving a step closer to cell dynamics, we consider that the ribosomes are distributed over small several cytoplasmic compartments inside the cell. Let $T$ represents this local pool of resource available in the immediate vicinity of the toggle switch. Thus our consideration of the two participants of the concerned toggle switch collects resource ribosome from the pool $T$ stands from its biological relevancy without any doubt. 
\\The available mRNA pool for translation is represented by $g_x$ and $g_y$ for protein $X$ and $Y$ respectively.
\\The available mRNA pool $g_x$ and $g_y$, collecting ribosomes from the pool $T$, with affinities $res_x$ and $res_y$ makes a ribosome bound complex $c_x$ and $c_y$ which will be translated to protein $X$ and $Y$ at a rate of $\epsilon_x$ and $\epsilon_y$ respectively. This asymmetry in resource allocation is very insightful here. Polycistronic mRNA pool in most of the bacterial organisms contains a multiple ribosome binding sites (RBS) \cite{burkhardt2017operon}, the rate of translation depends on the rate of recruitment of ribosomes to this RBS, as well as on the rate of translation initiation. The rate of ribosome recruitment also depends upon many factors. Including all this rates, we generalise the resource allocation or the protein production can have different rates as well. 
\\As mentioned, from the total pool $T$, the ribosome bound complex are presented by $c_x$ and $c_y$, thus further free pool of resource ribosome for translation is estimated by $(T-c_x-c_y)$.
\\The mutual repression is captured by Michaelis–Menten type term in our model and the hill function co-operativity $n$ is taken as $2$. The ODE representing the scenario is given by Eq. \ref{eqn tog}.
\begin{align}\label{eqn tog}
\frac{dc_x}{dt}=res_x\;(T-c_x-c_y)\;g_x-c_x\nonumber\\
\frac{dX}{dt}=\frac{c_x\;\epsilon_x}{1+y^n}-X\\
\frac{dc_y}{dt}=res_y(T-c_x-c_y)\;g_y-c_y\nonumber\\
\frac{dY}{dt}=\frac{c_y\;\epsilon_y}{1+x^n}-Y\nonumber
\end{align}
In steady state all the rate of changes is equal to $0$, we investigated the system.
\section{Result section}\label{sec 2}
\subsection{Pitchfork bifurcation in genetic toggle}
Pitchfork bifurcation occurs at specific equilibrium with perfect symmetry conditions of the toggle system. For one variable dynamical system, several studies are present while for a two variable dynamical system, a few studies in recent past \cite{bose2019bifurcation} investigated this phenomena in a brief.
For a conventional pitchfork model, we take the production rates of the two proteins to be equal, thus here we take $\epsilon_x$, the production rate of $X$ from its complex $c_x$ and, $\epsilon_y$, rate of production of $Y$ from its complex $c_y$ is equal. 
\subsubsection{Resource affinity regulates the symmetry of pitchfork bifurcation }
\begin{figure}
    \centering
    \includegraphics[width=16cm]{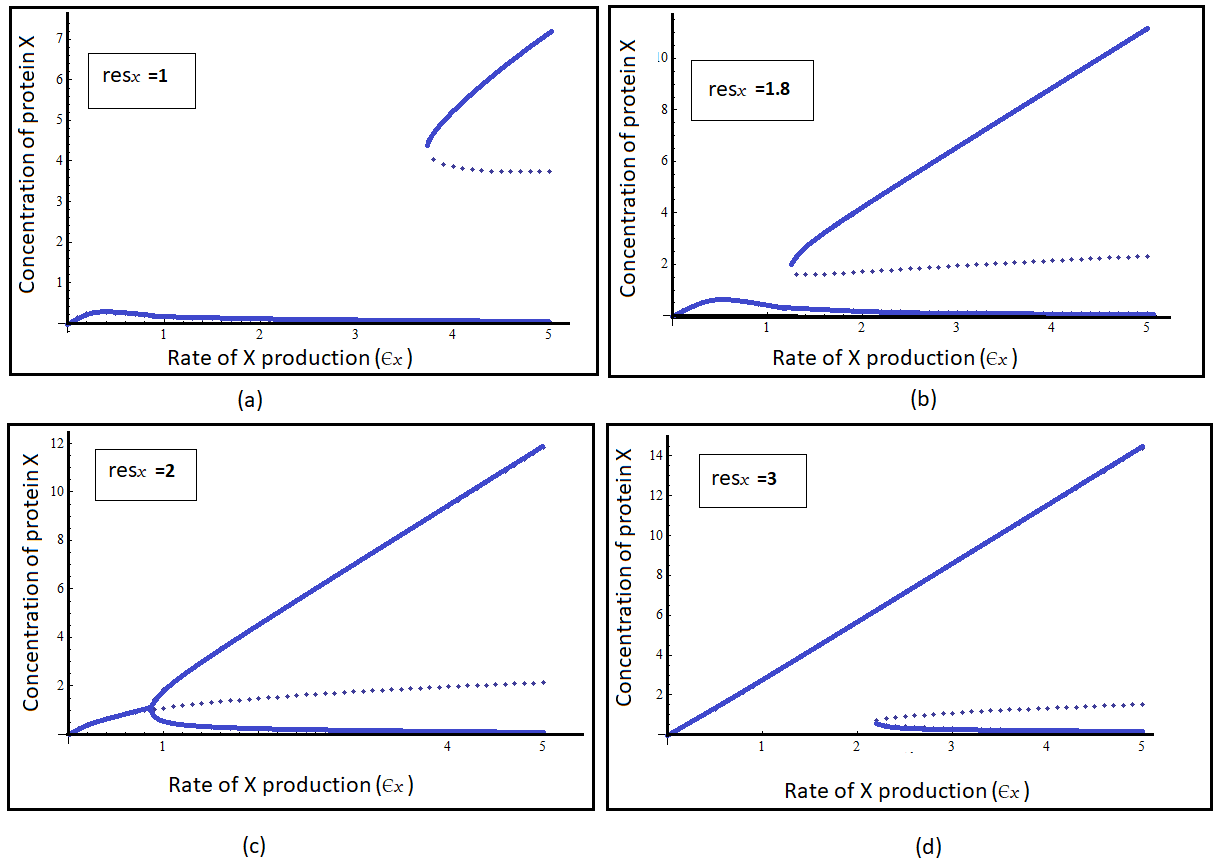}
    \caption{Resource affinity regulates the symmetry of pitchfork bifurcation. Concentration of protein $X$ wrt the rate of $X$ production $\epsilon_x$ plot. $n=2, res_y=2, g_x= g_y=5, T=5$ for all the plots. Resource affinity for $X$ production i.e. $res_x=1$ for (a), $res_x=1.8$ for (b), $res_x=2$ for (c), $res_x=3$ for (d).}
    \label{res affi}
\end{figure}

Following the conventional way of pitchfork bifurcation in toogle system, to the the protein production rates to be equal, a detailed literature review strongly supports our consideration of taking different resource affinity value without any loss of generality. We find resource affinity value regulates the symmetry of pitchfork bifurcation significantly. We investigated the model for fixed values of $n = 2, g_x = g_y = 5, T = 5, res_y = 2$ and changing $\epsilon_x$ such that for every point  $\epsilon_x = \epsilon_y$ for  three different values of $res_x$. When $res_x = res_y$ we find a beautiful symmetric pitchfork bifurcation in output while more interestingly asymmetry in resource affinity values destroys the symmetry in output pitchfork with a significant impression.  Starting with a lower resource affinity for $X$, $res_x < res_y$, a smooth transition of the pitchfork to low state of the system as shown in Fig. \ref{res affi}a, while higher state is only accessible for a very large perturbation in the system.  While starting from a higher resource affinity $res_x > res_y$ for Fig.\ref{res affi}d, the continuous accessible state is the higher production state, while lower production states are only accessible for only a large perturbation in the system as well. Same resource affinity of $X$ and $Y$ ($res_x= res_y=2$ for Fig.\ref{res affi}c) results in a symmetric pitchfork.  Also, it is interesting to note that, larger the asymmetry, higher the stability of the chosen transitioned state and bare the chances for its transition to another steady state even when perturbation is present from outside (comparing Fig.\ref{res affi}a, Fig.\ref{res affi}d with Fig.\ref{res affi}c). So, the results indicate to the conclusion that an initial asymmetry in resource allocation pre-patterns the cells to a higher production regime or in a lower production regimes and determines the cell fate.
\subsubsection{Availability of total mRNA pool regulates the symmetry of pitchfork bifurcation.}
\begin{figure}
    \centering
    \includegraphics[width=16cm]{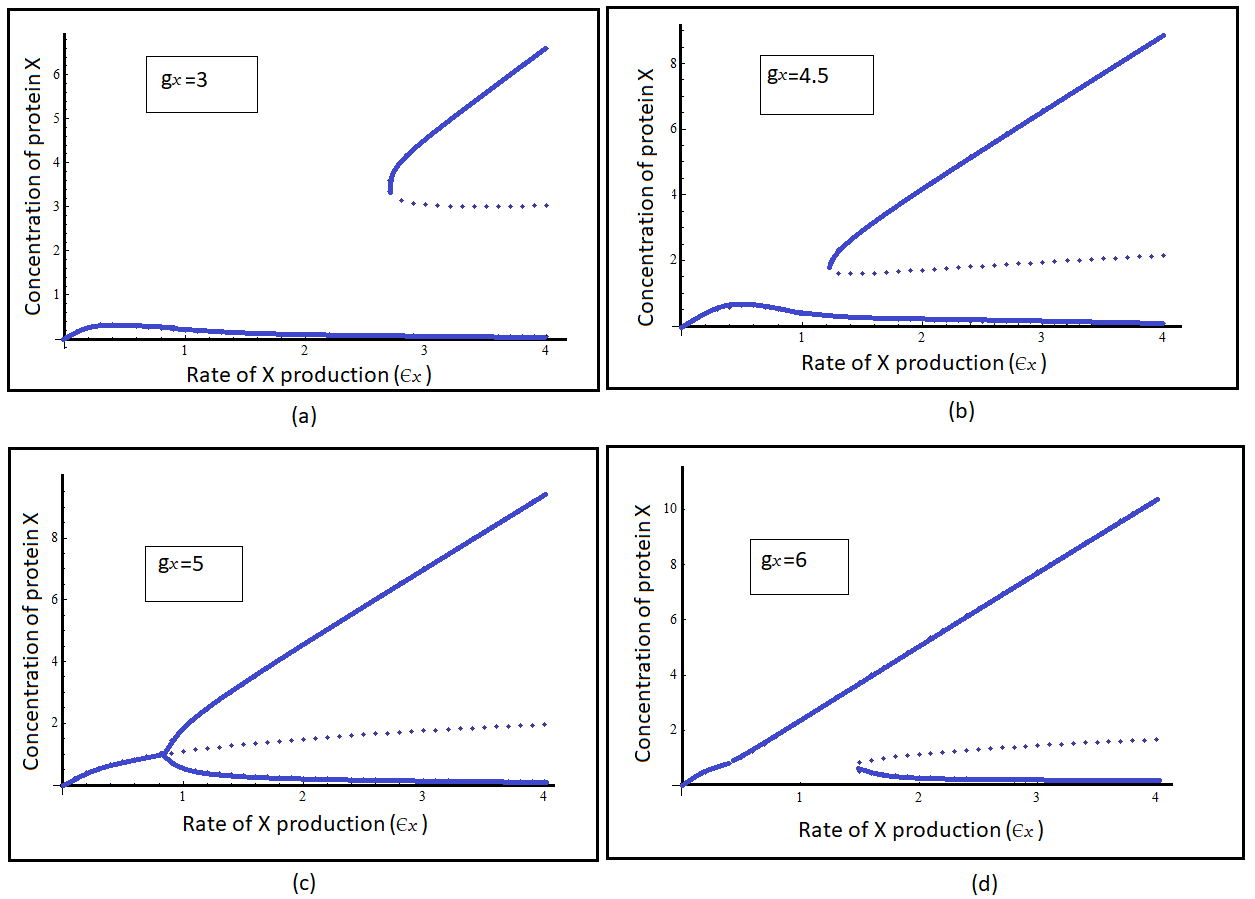}
    \caption{Availability of total mRNA pool regulates the symmetry of pitchfork bifurcation hence cell fate transition. Concentration of protein $X$ vs. rate of $X$ production $\epsilon_x$ plot. $res_x=res_y=2, n=2, T=5, g_y=5$ is fixed for all the plots and $\epsilon_x=\epsilon_y$. Gene copy numbers available for $X$ production i.e $g_x=3$ for (a), $g_x=4.5$ for (b), $g_x=5$ for (c), $g_x=6$ for (d).  }
    \label{pit gene copy}
\end{figure}

We find similar result wrt. gene copy number available for translation for the particular protein. The number of copies of a particular gene present in the genotype is usually called the gene copy number. A symmetry in presence of gene copy number with symmetry in other system parameters shows a perfect pitchfork in output while asymmetry in initial condition of gene copy number pre-patterns the system to a higher or lower production state according to available gene copy number for translation as shown in Fig.\ref{pit gene copy}. It is interesting to note that greater the asymmetry in initial gene copy number, greater the perturbation the system demands to transit from its continuous steady state to the other. 
\subsubsection{Total resource availability regulates the point of bifurcation in the system}
\begin{figure}
    \centering
    \includegraphics[width= 16cm]{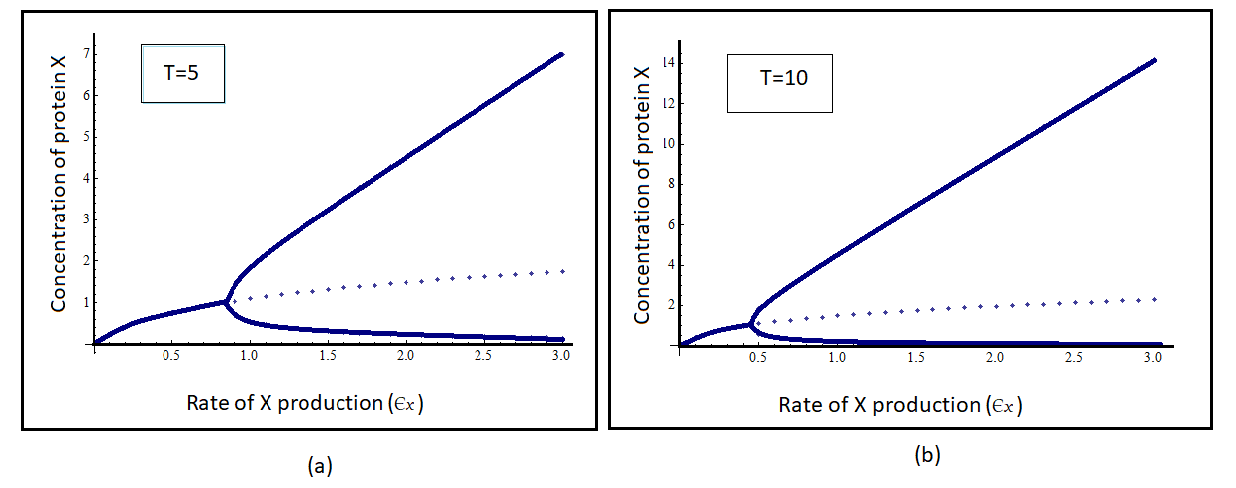}
    \caption{Total resource $T$ availability regulates the point of pitchfork bifurcation in genetic toggle. Concentration of protein $X$ vs. rate of $X$ production $\epsilon_x$ plot. $n=2, g_x=g_y=5, \epsilon_x=\epsilon_y, res_x=res_y=2$ for both the plot. $T=5$ for (a) plot, $T=10$ for (b) plot.  }
    \label{total T}
\end{figure}

We find, total resource availability significantly regulates the bifurcation point of the system as shown in Fig.\ref{total T}. We investigate the model for two fixed values of $T$, increasing the total resource available for translation the position of bifurcation comes to a lower value of input signal. Also the range of steady states drastically changes.
\subsection{Saddle node bifurcation in genetic toggle}
The genetic toggle, most conventionally known as genetic toggle switch, is biologically most important for its On/Off switch like behavior which plays a significant role in determining cell fate. From its early invention, researchers are deliberately searching for the ways of robust controlling of toggle switch. We find beautiful control on the toggle switch by controlling resource distribution.
\subsubsection{Resource distribution regulates the point of bifurcation in toggle switch}
We find the resource distribution significantly regulates the point of bifurcation in saddle node of genetic toggle as well, as shown in Fig.\ref{sd resx}. We investigate the system for $3$ different values of $res_x$ keeping all other parameters fixed at $res_y = 2, n = 2, g_x = g_y = 5, T = 5, \epsilon_y =2$ and plot concentration of protein $X$ wrt. activator of $X$ production $\epsilon_x$. Considering the blue line (continuous and dashed) primarily, which shows the scenario when $res_x=res_y=2$ that is resource allocation for $X$ to $Y$ is same, we find a change in $res_x$ shifts the curve left to the green curve (or right to the red curve) for a higher affinity for resource to $X$ than $Y$, $res_x = 3, > res_y = 2$ (for a lower affinity for resource to $X$ than $Y$, $res_x = 1, < res_y = 2$).
\subsubsection{The region of interest in toggle switch is significantly regulated by resource allocation}
\begin{figure}
    \centering
    \includegraphics[width =16cm]{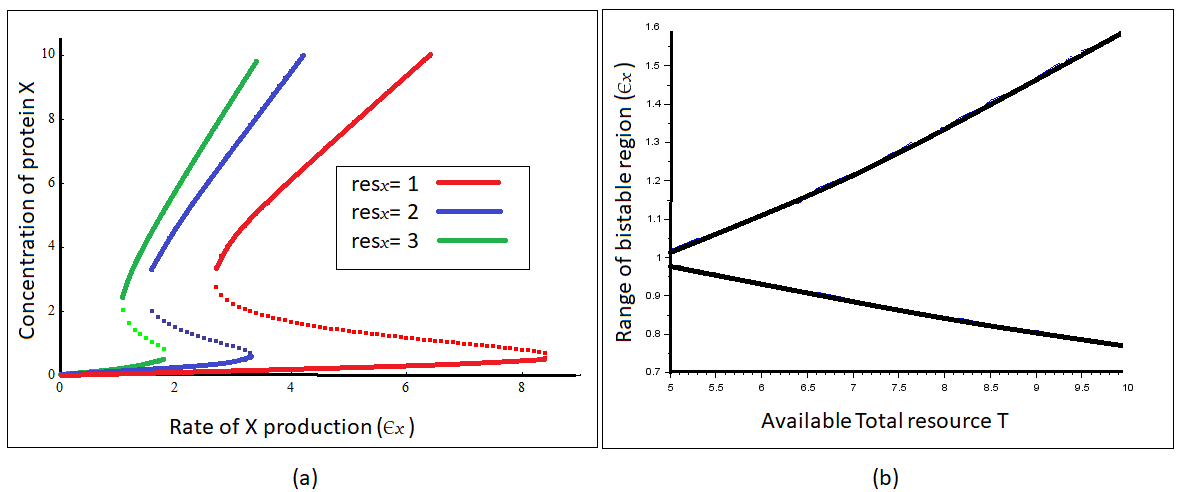}
    \caption{(a). Rate of $X$ production $\epsilon_x$ vs. concentration of protein $X$ plot in case of saddle node bifurcation in genetic toggle. $res_y=2, g_x=g_y=5, T=5, n=2 \epsilon_y=2$. (b). Phase plot of rate of $X$ production $\epsilon_x$ vs. total resource availability $T$. $res_x=res_y=1, n=2, g_x=g_y=5,\epsilon_y=1$}
    \label{sd resx}
\end{figure}

For a saddle node bifurcation, the most interesting region is the range of input signal for which the output protein concentration attains two drastically different concentration depending upon the mode of forward or backward operation. When investigated in bifurcation diagram, a set of stable equilibrium point is separated by unstable equilibrium points, the system cannot achieve physically. From Fig.\ref{sd resx} we get also the range of interest increases (or decreases) for $X$ with a lower resource allocation $res_x=1 < res_y=2$ (for $X$ getting higher resource than $Y$, $res_x=3 >res_y =2$). It is interesting to note that, lower resource availability to $X$, than $Y$, stabilises the bifurcation curve for larger fluctuation. This not only giving us an opportunity of robust controlling of the system, but also signifies initial biasing of system towards resource significantly modifies cell fate in terms of stability also. 
\subsubsection{Total resource availability T regulates Saddle node bifurcation curve}
Along with the variation with resource affinity, the availability of the total nutrients, say resource here significantly regulates the saddle node bifurcation curve as shown in Fig.\ref{sd resx}b. The regulation is quite positive here though. Greater resource stabilises the system for larger range of bistability, greater the switch robustness, while with less availability of resource switch response is not stable and the steady states can alter even for low fluctuation in the system.

\section{Conclusions and perspectives} \label{section:conclusion}
Cellular decision making is a fundamental biological phenomena by which a cell opts the different states prior to environmental conditions, leading to asymmetric cell differentiation. The underlying reason behind this, is still not entirely explored. We take a simple genetic motif, genetic toggle here, and shows that resource affinity asymmetry of the toggle participants, both for the saddle node and in pitchfork bifurcation significantly biases the cell fate. This mutually repressing motif is very common in nature \cite{saka2007mechanism, clark2017dynamic} where in output the system shows patterning by choosing one cell fate over other. The availability of total resource pool significantly regulates the bifurcation point in the motif. While in case of synthetic circuit, this resource limitation is very true because the gene circuit implemented in host, entirely depends upon host's resource for its expression, in case of cell the limitation in cellular resource ribosome is major factor, indicating our findings to be true also. We also investigated the effect of gene copy number in case of pitchfork bifurcation indicating an initial asymmetry biases the cell fate to lower or higher production states accordingly.
\\ Here, it is important to note that our entire consideration is valid for low growth state of the system. The effect of growth rate on cell dynamics is well  established  \cite{tan2009emergent}. Over expression of endogenous genes, or adding some synthetic construct in cell, destabilises the resource distribution, makes the growth rate smaller, while growth causing dilution enhances the protein degradation. So, some researchers point out that growth is a significantly regulatory parameter in every cellular phenomena. But some experimental results also  pointed out that these effects only depends upon experimental conditions, causing some momentary changes in dynamics. Our study mostly follows the experimental situation \cite{gyorgy2015isocost,shachrai2010cost} when the growth rate is low and competition effect is significant.
\\A perfect noise free environment in cell is impracticable. For a single cell, though the consideration does not violate the reality, working with a group of cell, the predictability can vary significantly. Addition of noise in existing dynamics will give a result close to reality. Also the limitation of other cellular nutrients in the way of gene expression can regulate the alternate protein production, stabilizing one state over other, regulating cell fate. In future, we would like to extend our work for a complete scenario including transcriptional, translational and degradation machinery competition in a noisy environment of cell.
\section*{Acknowledgement}
\noindent PC and SG acknowledge the support  by DST-INSPIRE, India, vide sanction Letter No. 
\\DST/INSPIRE/04/2017/002765 dated- 13.03.2019.
\bibliographystyle{abbrv}
\bibliography{Bibliography.bib}

\end{document}